\documentclass[%
 aip,
 amsmath,amssymb,
 reprint,%
]{revtex4-1}

\usepackage{graphicx}
\usepackage{dcolumn}
\usepackage{bm}
\usepackage{graphicx}
\usepackage{float}
\usepackage[utf8]{inputenc}
\usepackage[T1]{fontenc}
\usepackage{mathptmx}
\usepackage{etoolbox}

\makeatletter
\def\@email#1#2{%
 \endgroup
 \patchcmd{\titleblock@produce}
  {\frontmatter@RRAPformat}
  {\frontmatter@RRAPformat{\produce@RRAP{*#1\href{mailto:#2}{#2}}}\frontmatter@RRAPformat}
  {}{}
}%
\makeatother
\begin{document}

\preprint{AIP/123-QED}

\title[Sample title]{Progress toward a  microwave frequency standard based on  laser-cooled large scale ${}^{{\rm{171}}}{\rm{Y}}{{\rm{b}}^{\rm{ + }}}$ ion crystal  }
\author{N.C.Xin}
\affiliation{ 
State Key Laboratory of Precision Measurement Technology and Instruments, Key Laboratory of Photon Measurement and Control Technology of Ministry of Education, Department of Precision Instrument, Tsinghua University, Beijing 100084, China
}%

\author{H.R.Qin}

\affiliation{ 
State Key Laboratory of Precision Measurement Technology and Instruments, Key Laboratory of Photon Measurement and Control Technology of Ministry of Education, Department of Precision Instrument, Tsinghua University, Beijing 100084, China
}%
\affiliation{%
Department of Physics, Tsinghua University, Beijing 100084, China
}%

\author{S.N.Miao}
\affiliation{ 
State Key Laboratory of Precision Measurement Technology and Instruments, Key Laboratory of Photon Measurement and Control Technology of Ministry of Education, Department of Precision Instrument, Tsinghua University, Beijing 100084, China
}%
\author{Y.T.Chen}
\author{J.Z.Han}
\affiliation{ 
State Key Laboratory of Precision Measurement Technology and Instruments, Key Laboratory of Photon Measurement and Control Technology of Ministry of Education, Department of Precision Instrument, Tsinghua University, Beijing 100084, China
}%


\author{J.W.Zhang}%
 \homepage{http://faculty.dpi.tsinghua.edu.cn/team/zhangjw}
 \email{zhangjw@tsinghua.edu.cn}
\affiliation{ 
State Key Laboratory of Precision Measurement Technology and Instruments, Key Laboratory of Photon Measurement and Control Technology of Ministry of Education, Department of Precision Instrument, Tsinghua University, Beijing 100084, China
}%

\author{L.J.Wang}
 \affiliation{ 
State Key Laboratory of Precision Measurement Technology and Instruments, Key Laboratory of Photon Measurement and Control Technology of Ministry of Education, Department of Precision Instrument, Tsinghua University, Beijing 100084, China
}%
\affiliation{%
Department of Physics, Tsinghua University, Beijing 100084, China
}%

\begin{abstract}

We report on progress towards  a  microwave frequency stand ard based on a laser-cooled ${}^{{\rm{171}}}{\rm{Y}}{{\rm{b}}^{\rm{ + }}}$ ion trap system. The electronics , lasers, and magnetic shields are integrated into a single physical package. With over $10^5$ ions are  stably trapped, the system offers a high signal-to-noise ratio Ramsey line-shape. In  comparison with previous work, the frequency instability of a ${}^{{\rm{171}}}{\rm{Y}}{{\rm{b}}^{\rm{ + }}}$ microwave clock was further improved to  c for averaging times between 10 and 1000 s.

\end{abstract}

\maketitle

%

In recent years,  numerous atomic clocks have been developed that offer precise and stable time frequency signals. Such signals are crucial for both basic physics \cite{dzuba2016strongly,safronova2018search} and practical applications\cite{hinkley2013}.  In particular, with the excellent performance, atomic clocks operating with a cloud of trapped ions are showing extremely strong promise  in the quest for high precision. For instance,  optical clocks based on $\rm{Al}^+$ exhibit a systematic uncertainty\cite{brewer2019al+} of  $9.4 \times { 10^{- 19}}$. Whereas   $\rm{Hg}^+$ microwave clock \cite{liu2020progress} demonstrate a frequency instability of $1.5\times {10^{ - 13}}/\sqrt \tau$\cite{burt2021demonstration}. In targeting the microwave spectral region,  $\rm{Hg}^+$\cite{burt2021demonstration,berkeland1998laser}, $\rm{Cd}^+$\cite{miao2021precision,miao2015high}, $\rm{Yb}^+$\cite{mulholland2019compact,king2012absolute}, and $\rm{Ba}^+$\cite{pino2020demonstration} trap systems have  been  investigated meticulously during the past few decades. The investigations show that while ensuring a decent performance, an atomic microwave frequency standard based on trapped ions has great potential in high transportability  because their  interaction time are long and the   structure of the system is relative simple. For these reasons, microwave ion clocks are being touted as  promising candidates for the next generation of practical atomic clocks. 

Much progress on microwave clocks based on ytterbium ions (${}^{{\rm{171}}}{\rm{Y}}{{\rm{b}}^{\rm{ + }}}$)  has been achieved. For the buffer gas cooled $\rm{Yb}^+$ microwave clock,  an instability of $10^{-11}/\sqrt \tau $ has been demonstrated, which was limited by collision and pressure related shifts. The instability of the laser-cooled $\rm{Yb}^+$ microwave clock  was measured to be  $2.09 \times {10^{ - 12}}/\sqrt \tau$ and was limited by second-order Zeeman  effects\cite{phoonthong2014determination}. In absolute terms, the frequency instability potential of a laser-cooled $\rm{Yb}^+$ microwave clock is the same as for the microwave clocks based on $\rm{Hg}^+$ and $\rm{Cd}^+$\cite{jau2012low,schwindt2016highly,warrington1999csiro,schwindt2018operating}.  Lately, researchers at the National Physical Laboratory (NPL) have built a prototype of a   laser-cooled  ${}^{{\rm{171}}}{\rm{Y}}{{\rm{b}}^{\rm{ + }}}$ microwave clock that is both compact and portable.  The entire system  fits into a 6U 19-inch rack unit $(51\times 49\times 28)$ $\rm{cm}^3$, verifying the feasibility of such miniaturized Ytterbium-based clocks.\cite{mulholland2019laser}.


In this study, we report the design and prototyping  of a  laser-cooled ${}^{{\rm{171}}}{\rm{Y}}{{\rm{b}}^{\rm{ + }}}$ microwave clock, based on the ground-state hyperfine transition of ${}^{{\rm{171}}}{\rm{Y}}{{\rm{b}}^{\rm{ + }}}$  at 12.6 GHz. The clock system  is well integrated into a single physical package and  operates successfully in the laboratory environment. More than $10^5$  ytterbium ions are stably trapped  in the system enabling it to offer a high signal-to-noise ratio (SNR) Ramsey signal of the ground-state transition. The short-term stability of the ${}^{{\rm{171}}}{\rm{Y}}{{\rm{b}}^{\rm{ + }}}$ microwave clock has been improved to $8.5 \times {10^{ - 13}}/\sqrt \tau$,  a current record level, and its performance in practice demonstrates that the laser-cooled ${}^{{\rm{171}}}{\rm{Y}}{{\rm{b}}^{\rm{ + }}}$ clock has promising potential that is as good as other types of ion trap systems. Our research is helpful for developing a compact microwave clock device that can used in ground-based time-keeping, navigation station networking, and terrestrial network synchronizing.


\begin{figure}[H]
\centering
\includegraphics[width=8cm]{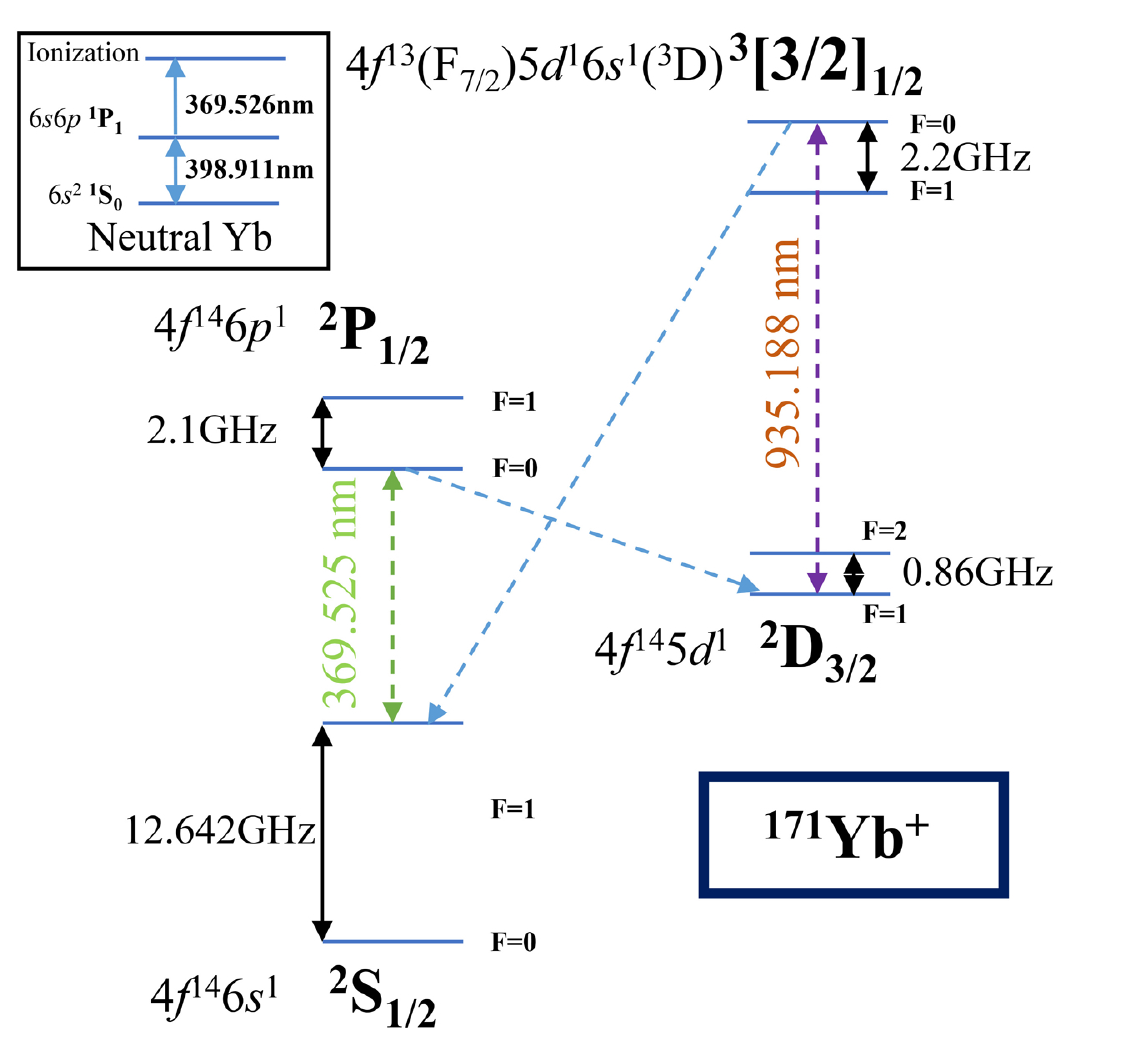} 
\caption{\label{fig:epsart} (Color online) Schematic energy levels of single ${}^{{\rm{171}}}{\rm{Y}}{{\rm{b}}^{\rm{ + }}}$ ion (not to scale).}
\end{figure}

The specifics of the operation of an microwave clock are determined by the electronic energy-level structure of the trapped ions. The simple energy-level structure of single ${}^{{\rm{171}}}{\rm{Y}}{{\rm{b}}^{\rm{ + }}}$ is shown in Fig.1. To cool the ions and probe their states, a 369 nm laser is used to cycle between the transition states labeled  ${}^2{S_{1/2}}(F = 1)$ and ${}^2{P_{1/2}}(F = 0)$. The neutral Yb atoms are ionized using a combination of two laser beams of wavelengths 369 nm and 399 nm. An extra 935 nm laser is used to drive ions out of the long-lived state ${}^2{D_{3/2}}(F = 0)$\cite{yu2000lifetime}. Moreover,  to drive ions out of the state ${}^2{S_{1/2}}(F = 1)$ during cooling, microwave radiation of 12.6 GHz is also applied. The ground state hyperfine splitting, $({}^2{S_{1/2}}(F = 1) \to {\rm{ }}{}^2{S_{1/2}}(F = 0))$ of 12.6 GHz, is used as the clock transition.

\begin{figure}[H]
\includegraphics[width=8cm]{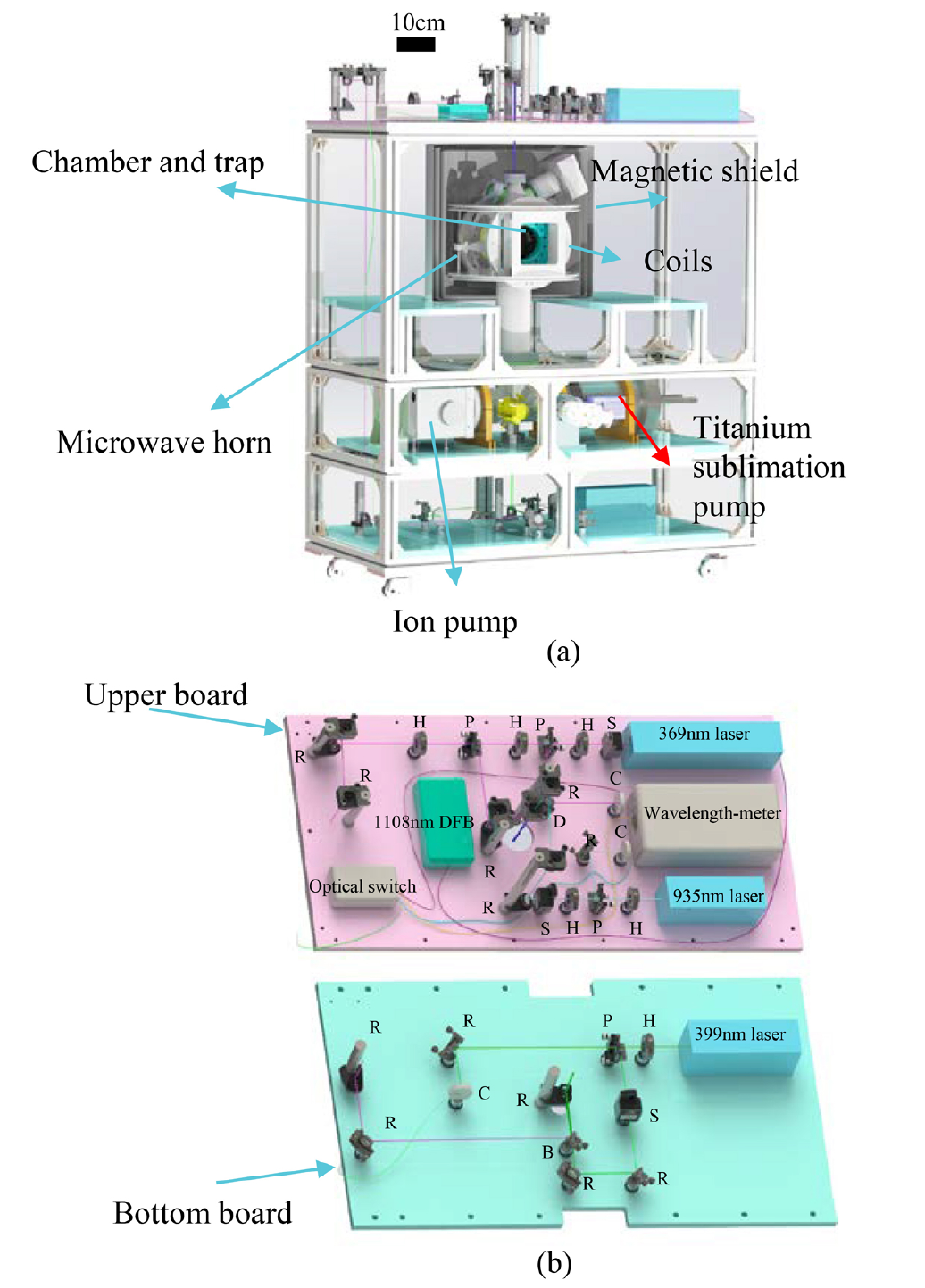}
\caption{\label{fig:wide} Detailed sketch of the  ${}^{{\rm{171}}}{\rm{Y}}{{\rm{b}}^{\rm{ + }}}$ ion  microwave clock system. (a)
System overview: Upper board : lasers of 369 nm and 935 nm for cooling ${}^{{\rm{171}}}{\rm{Y}}{{\rm{b}}^{\rm{ + }}}$; middle layer: vacuum chamber, linear Paul trap, magnetic shield, magnetic coils, microwave horn, ion pump , and titanium sublimation pump ; bottom board: lasers of 369 nm and 399 nm for ionizing ${}^{{\rm{171}}}{\rm{Y}}{{\rm{b}}}$.  (b)  Detailed information of laser system. H: half-wave plate; P: polarization beam splitter; D: dichromatic mirror; S: optical shutter; B: beam splitter; C: fiber-optic couplers; R: Reflection mirror; The cabinet with photomultiplier tube, microwave synthesizer,  electron-multiplying charge-coupled device (EMCCD) , and hollow cathode lamp (HCL) are not shown.}
\end{figure}


In regard to the clock system (Fig. 2), an  ${}^{{\rm{171}}}{\rm{Y}}{{\rm{b}}^{\rm{ + }}}$ cloud  is trapped in a well-designed linear Paul trap. Details of the ion trap are given elsewhere \cite{han2021toward}, and hence only a brief description is presented here. The trap is made up of four trisected cylindrical electrodes made of copper. The radius of each electrode is ${r_e} = 7.1$ mm, and the shortest distance between the electrodes and the ion trap center is ${r_o} = 6.2$ mm.  The ratio of
${\raise0.7ex\hbox{${{r_e}}$} \!\mathord{\left/
 {\vphantom {{{r_e}} {{r_o}}}}\right.\kern-\nulldelimiterspace}
\!\lower0.7ex\hbox{${{r_o}}$}}$ is optimized and set at 1.15, to reduce RF heating effects and to increase the number of trapped ions\cite{denison1971operating}.
The lengths of the electrode sections are 20, 40, and 20 mm. Direct current (DC) voltage is applied across adjacent sections, whereas a radio frequency (RF) voltage, with frequency fixed at 1.144 MHz, is assigned to the central section. Usually, the amplitude of the RF voltage is set below 200 V, implying that the trapping $q$ parameter is smaller than 0.11. A DC voltage of 4 V is adopted to provide fine axial confinement of the ion cloud. Besides the ion pump, a titanium sublimation pump is added to enhance the vacuum degree because  ${}^{{\rm{171}}}{\rm{Y}}{{\rm{b}}^{\rm{ + }}}$ is found to bond rapidly with some molecules such as  ${{\rm{O}}_2}$, ${{\rm{H}}_2}$, and ${{\rm{H}}_{\rm{2}}}{\rm{O}}$\cite{sugiyama1995disappearance,hoang2020ybh+} [see Fig. 2(a)].
A background pressure below ${10^{ - 9}}$ Pa is maintained during clock operations. For the Helmholtz coil system, three orthogonal pairs of coils are arranged compactly around the vacuum chamber and provide a steerable magnetic field that compensates the residual magnetic field and performs the Zeeman sublevel splitting. An extra ${\rm{700 }}$ ${\rm{\mu T}}$ external magnetic field, generated by a pair of coils, is also placed in the north-south direction to induce a precession in the dipole moments of the ions and to destabilize the dark states\cite{berkeland2002destabilization}.To minimize the second-order Zeeman (SOZS) shift, a magnetic shield, composed of three layers Permalloy, wraps the vacuum chamber and coils.  The shielding factors in the north-south, east-west, and up-down directions are measured to be 1000, 2000, and 300, respectively.

In regard to the laser system setup [Fig. 2(b)], all wavelengths of the various laser beams, corresponding to the different state transitions, were found by using saturated absorption spectroscopy that employed a Yb hollow cathode lamp. For accuracy, a high-resolution wavemeter (HighFinesse WS8-2) together with a proportional--integral--derivative controller was used to measure and stabilize the lasers. The 399 nm and 935 nm lasers are respectively extended-cavity-diode and distributed-feedback (DFB) lasers. Because the 369 nm laser beam plays a pivotal role in both cooling and probing, to enhance the aseismic capacity and large mode-hop-free tuning range, this beam is generated by a frequency-tripled laser---the 1108 nm DFB fiber laser---and two periodically poled crystals. Reviewing the laser system, the 369 nm laser beam is separated into two beams by a polarization beam splitter. One of the beams is combined by a dichromatic mirror with the 935 nm laser beam and passed from above into the vacuum chamber. The other beam is combined with the 399 nm laser beam and directed in the opposite direction to the 935 nm laser beam. Because the extra strong magnetic field is recommended to be at a certain oblique angle relative to the polarization of the laser beams, all beams are linearly polarized by half-wave plates. Our lasers can provide power outputs up to 70 mW, 100 mW, and 80 mW for the 369 nm, 399 nm, and 935 nm beams, respectively, propagating in free space. The spatial configuration of the ion cloud is captured by an  electron-multiplying charge-coupled device (EMCCD). A photomultiplier tube (PMT) is used to record the number of photons emitted by the ions. The microwave radiation is generated by a commercial microwave synthesizer (Agilent E8257D), referenced by the 10 MHz signal obtained from an oven-controlled crystal oscillator (OCXO).

\section{\label{sec:level1}Experiment result}
To load the ions, the 369 nm laser is detuned by approximately 50 MHz to the low-frequency side of the corresponding transition. Because the natural Yb solid metal used is a mixture of various stable isotopes, the 399 nm laser is tuned to the peak-frequency of  the ${}^1{S_0} \to {\rm{ }}{}^1{P_1}$ transition  of ${}^{{\rm{171}}}{\rm{Y}}{{\rm{b}}}$ atom to achieve isotope selection. Confined by both the RF and DC voltages, a long narrow  ${}^{{\rm{171}}}{\rm{Y}}{{\rm{b}}^{\rm{ + }}}$ cloud is obtained and captured by the CCD camera (Fig. 3).


\begin{figure}[H]
\centering
\includegraphics[width=8cm]{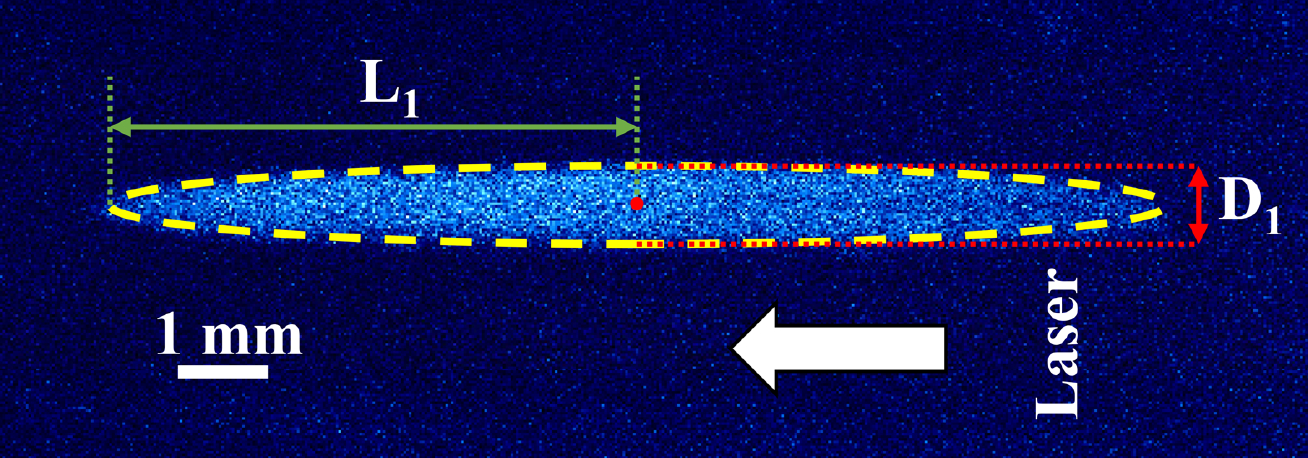} 
\caption{\label{fig:epsart} EMCCD image of a trapped ${}^{{\rm{171}}}{\rm{Y}}{{\rm{b}}^{\rm{ + }}}$ ion cloud. The semi-major and semi-minor axes of the ellipsoid are ${{\rm{L}}_1} = 6.07$ ${\rm{ mm}}$ mm and $\frac{{{{\rm{D}}_1}}}{2} = 0.48$ ${\rm{ mm}}$. The amplitude of the RF voltage is approximately 200 V; the DC voltage is set to 4 V. The image has been rotated ${90^\circ }$ for convenience; the long axis of the ion cloud is actually oriented vertically .
}
\end{figure}

To estimate the number of ions of a cloud, the number densities of ions are estimated by using low-temperature ion-density theory, from which we have\cite{hornekaer2001structural}:
\begin{eqnarray}
{n=\frac{\varepsilon_{0} V_{R F}^{2}}{M \Omega^{2} r_{o}^{4}}}
\end{eqnarray}
where ${\varepsilon _0}$ denotes the vacuum permittivity , M is the mass of  ${}^{{\rm{171}}}{\rm{Y}}{{\rm{b}}^{\rm{ + }}}$. ${V_{RF}} = 200$ V, $\Omega {\rm{ = 2\pi }} \times 1.144$ MHz
, and ${r_o} = 6.2$ mm.  The number density of ${}^{{\rm{171}}}{\rm{Y}}{{\rm{b}}^{\rm{ + }}}$ ions is estimated to be approximately ${n_{{\rm{Y}}{{\rm{b}}^ + }}} = 1.634 \times {10^{13}}$ $\rm{m^{-3}}$, the number of trapped ions being ${N_{{\rm{Y}}{{\rm{b}}^ + }}} = 1.1(0.3) \times {10^5}$.


To obtain an unambiguous Ramsey signal of the microwave clock transition at 12.6 GHz, all lasers, microwave radiation, and magnetic fields of the coils must be precisely controlled. A field-programmable gate-array circuit board is used to generate control signals to within millisecond order. During measurements, all laser frequencies are first stabilized to within 2 MHz by locking to the wavelength-meter. The ions are then trapped and cooled using the 369 nm and 935 nm laser beams directed downward into the system. A strong external magnetic field and high-power microwave radiation is applied to suppress the dark states.
 To pump the ions into the $({\rm{ }}{S_{1/2}}(F = 0))$ state via the decay from the off-resonance excited  $({\rm{ }}{}^2{P_{1/2}}(F = 1))$, the high-power microwave radiation is turned off. During the interaction with the 
$\pi /2$ pulse, only the steerable magnetic field is applied to compensate the residual magnetic field and splitting of the Zeeman sublevels. Finally, the upward-directed low-power 369 nm laser beam is used to detect the number of ions in the  $({\rm{ }}{}^2{S_{1/2}}(F = 1))$ states. An interval of approximately 20 ms is set to wait for closure and establishment of the strong magnetic field, the control of which is based on a transistor. From the Ramsey fringes (Fig. 4) recorded by the PMT, peak-to-peak values are calculated along with their amplitudes at half-waist, from which estimates of the SNR, typically 35, for the clock resonance are obtained. The high SNR of the Ramsey signal ensures a decent stability performance for the designed ${}^{{\rm{171}}}{\rm{Y}}{{\rm{b}}^{\rm{ + }}}$ microwave frequency standard.

\begin{figure}[H]
\centering
\includegraphics[width=8cm]{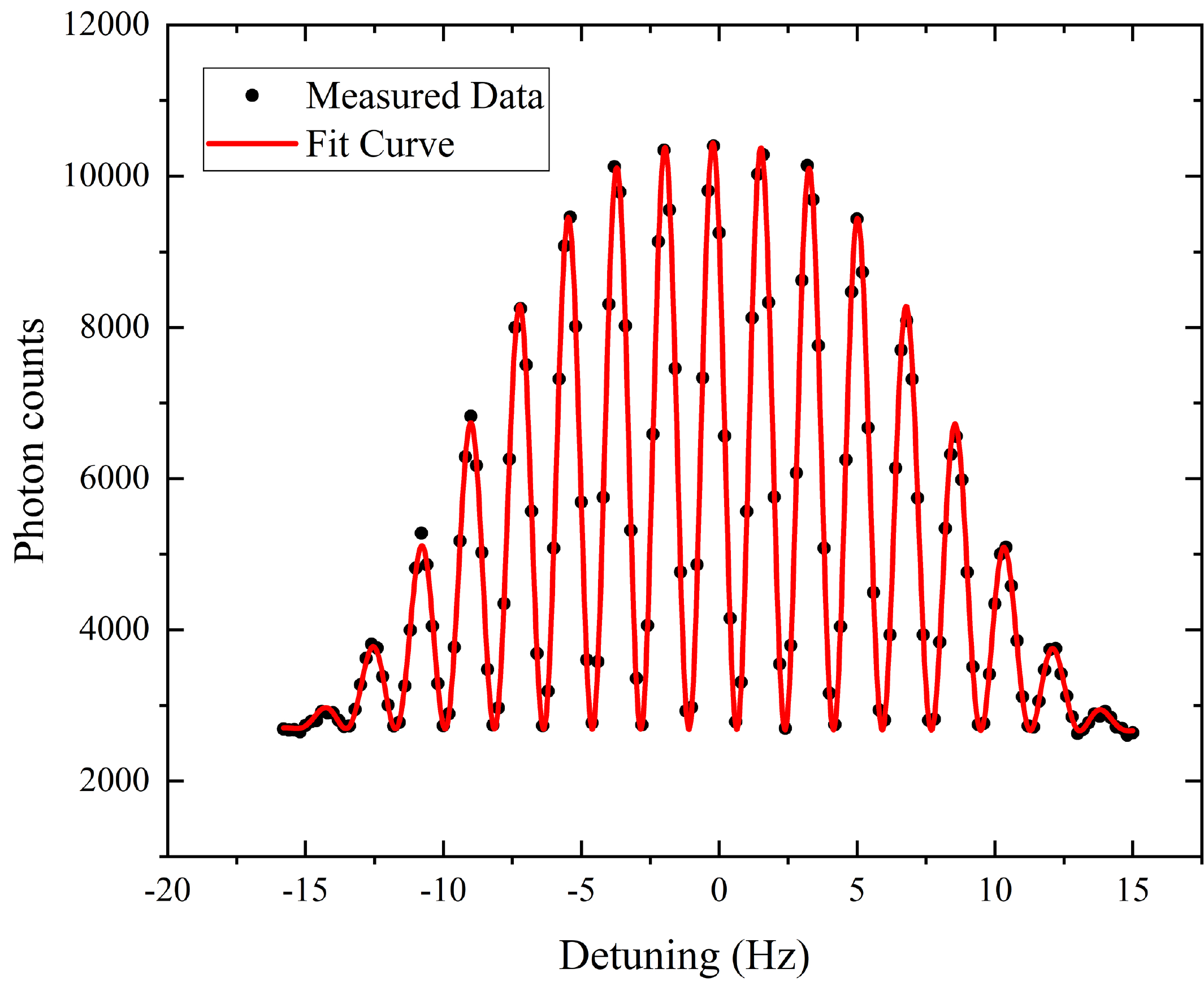} \\[5pt]  
\caption{\label{fig:wide}  Typical Ramsey lineshape of the clock transition (12.6 GHz) with a microwave pulse time of 60 ms, free time of 500 ms, microwave power of -29.8 dBm, and a fluorescence signal
integration time of 150 ms. Each data point is the average of five results. The fitted curve has a central frequency of 12642812121.47 Hz }
\end{figure}




The short-term frequency instability limitation can be  estimated after acquiring the Ramsey signal. Here, we have mainly considered quantum projection noise, pump noise, and shot noise, which are typical limitations to the frequency stability of a passive frequency standard\cite{itano1993quantum}. 

To estimate the quantum projection noise ${\sigma _{{\rm{proj}}}}$, pump noise ${\sigma _{{\rm{pump}}}}$ and shot noise ${\sigma _{{\rm{shot}}}}$, the maximum photon counts of the Ramsey signal ${S_{\max }}$,  the minimum  photon counts of the Ramsey signal ${S_{\min }}$,  the photon counts arising from background scattered light ${{S_{{\rm{back}}}}}$, and  the number of ions $n$   need to be confirmed first. Typically, for our system,  ${S_{\max }} = 10500$, ${\rm{ }}{S_{\min }} = 2500$, ${\rm{ }}{S_{{\rm{back}}}} = 1500$, ${\rm{ }}n = 100000$. Then the noise is calculated from\cite{itano1993quantum}:
 \begin{eqnarray}
{\sigma _{{\rm{proj}}}} = \sqrt {\eta n} \frac{K}{2}\\
{\sigma _{{\rm{pump}}}} = \sqrt {n\eta (1 - \eta )} \frac{K}{2}\\
{\sigma _{{\rm{shot}}}} = \sqrt {\frac{{{S_{\max }} + {S_{\min}}}}{2}} 
 \end{eqnarray}
 where $K = \frac{{({S_{\max }} - {S_{{\rm{back}}}})}}{n},\eta  = 1 - \frac{{{S_{\min }} - {S_{{\rm{back}}}}}}{{{S_{\max }} - {S_{{\rm{back}}}}}}$. Therefore, the overall calculated noise is:
  \begin{eqnarray}
{\sigma _{{\rm{cal}}}} = \sqrt {{\sigma _{{\rm{proj}}}}^2 + {\sigma _{{\rm{pump}}}}^2 + {\sigma _{{\rm{shot}}}}^2} 
 \end{eqnarray}
 
 With  ${\sigma _{{\rm{cal}}}} = 81$, the corresponding SNR is around 48.8 , which indicates that the frequency instability limitation is around $5.9 \times {10^{ - 13}}$ at 1 s.

We conducted a preliminary closed-loop run to verify the potential of the system. A detailed description of the measurement setup has been fully discussed\cite{zhang2014toward}. We set the free time at 1 s and the cooling time at 1.2 s for measurements during the closed-loop run to reduce the linewidth and increase the theoretical instability limitation. After comparing our 10 MHz output signal with  an active hydrogen maser the Allan deviations of the  frequency standard are presented (Fig. 5). Moreover, the frequency stability of our free-running OCXO, together with the performance of recent  microwave clocks of other research groups, are also presented for direct comparison\cite{phoonthong2014determination,mulholland2019compact}. The measured frequency stability of our clock is $8.5 \times {10^{ - 13}}/\sqrt \tau$ for averaging times between 10  and 1000 s.

 \begin{figure}[H]
\centering
\includegraphics[width=8cm]{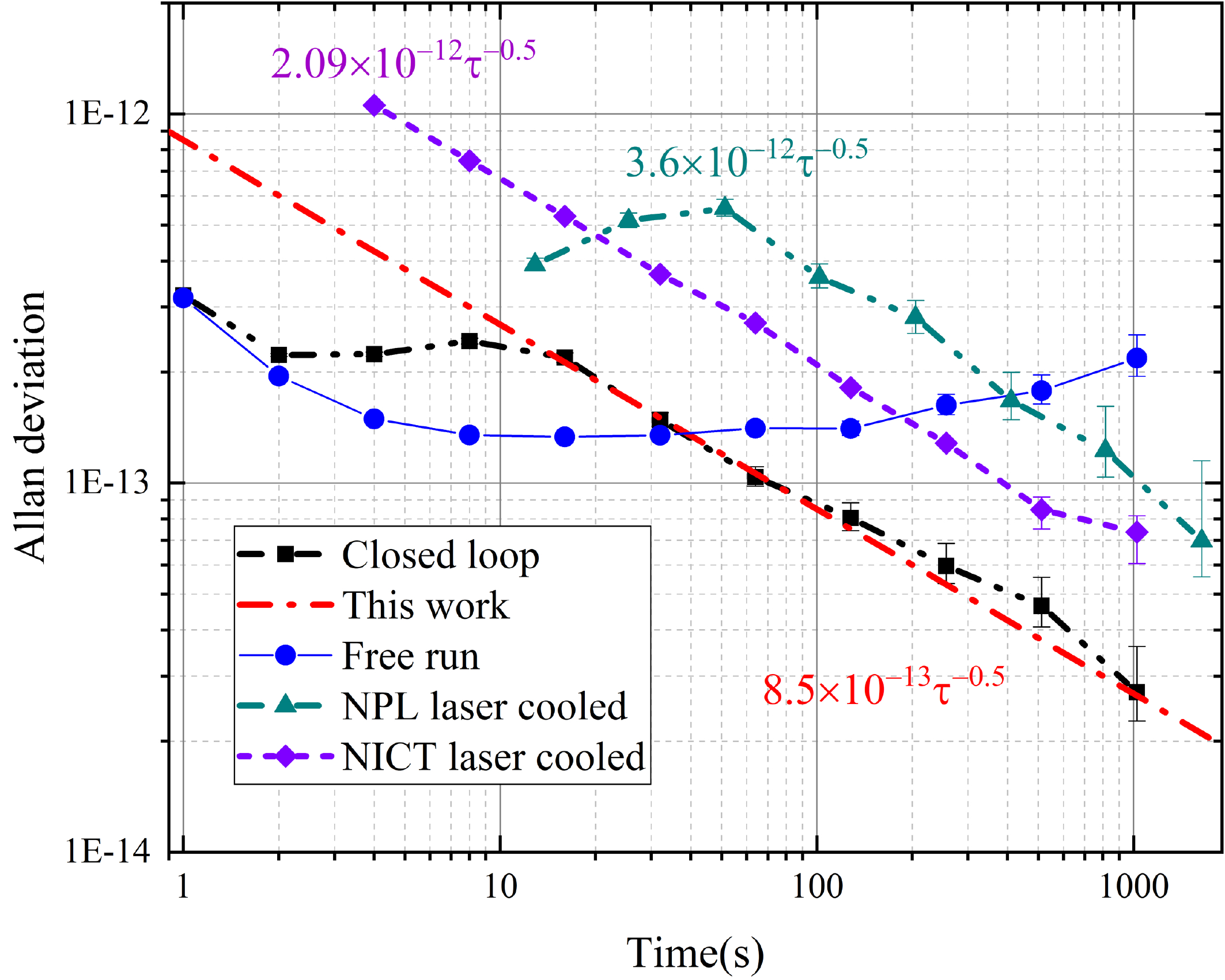} \\[5pt]  
\caption{\label{fig:wide}   Allan deviations of the ${}^{{\rm{171}}}{\rm{Y}}{{\rm{b}}^{\rm{ + }}}$ frequency standard
and the free-running OCXO. Other scholars' primary and significant work are also plotted, including results from National Institute of Information and Communications (NICT) and National Physical Laboratory (NPL)\cite{phoonthong2014determination,mulholland2019compact}. The system was run continuously for 13000s. The frequency reference during the
measurements is generated by a hydrogen maser. }
\end{figure}


Overall, as the number of trapped ions increases to around ${10^5}$, the SNR of our system is vastly enhanced, which explains the reason for the improved performance of our system. However, our system has not reached the performance limit that we expected from the calculation results. Fluctuations in the laser power, laser frequency, magnetic field, and working temperature are suspected in increasing noise and degrading system performance \cite{enzer2016drifts}. The slight bump in our measured Allan deviation at around 500 s (Fig. 5) is considered to be related to fluctuations in room temperature and laser power, the fluctuation period of which lasts thousands of seconds. To further enhance the short-term instability of our system and provide long-term system accuracy and stability, active feedback must be used to suppress these fluctuations, which is our objective for the next round of measurements.

We reported on progress towards a highly stable and high performance ${}^{{\rm{171}}}{\rm{Y}}{{\rm{b}}^{\rm{ + }}}$  microwave frequency standard.  In attaining  this microwave frequency standard, cooling, pumping and detecting of  ${}^{{\rm{171}}}{\rm{Y}}{{\rm{b}}^{\rm{ + }}}$ are achieved. In our experiment,  ${10^5}$ Yb ions are stably trapped. A high SNR Ramsey fringe line-shape is obtained and  the microwave frequency standard shows a short-term stability of $8.5 \times {10^{ - 13}}/\sqrt \tau$.

\begin{acknowledgments}
The research is supported by National Natural Science Foundation of China (12073015); National Key R$\&$D Program of China (2016YFA0302101); Beijing Natural Science Foundation (1202011); Tsinghua University Initiative Scientific Research Program.

We would like to thank C. F. Wu, L. M. Guo, T. G. Zhao, and W. X. Shi for reasonable advice and helpful suggestions.

\end{acknowledgments}

\section*{Data Availability Statement}

The data that support the findings of this study are available
from the corresponding author upon reasonable request.

\end{document}